\documentclass[10pt,conference]{IEEEtran}

\usepackage[pdftex]{graphicx}
\usepackage[usenames,dvipsnames,table,xcdraw]{xcolor}
\usepackage{array}
\usepackage{multirow}
\usepackage[normalem]{ulem}
\useunder{\uline}{\ul}{}
\usepackage{booktabs}
\usepackage{supertabular,booktabs}
\usepackage{longtable}
\usepackage{lscape}
\usepackage{cite}
\usepackage{url}
\usepackage{hyperref}
\usepackage{dirtytalk}
\usepackage{graphicx}
\usepackage{comment}
\usepackage{makecell}
\usepackage{svg}
\usepackage{amsmath}

\usepackage[caption=false]{subfig}
\graphicspath{{Images/}}
\usepackage[normalem]{ulem}
\useunder{\uline}{\ul}{}

\usepackage{tabularx}
\usepackage{balance}
\usepackage{nameref}
\usepackage{graphicx}

\usepackage[most]{tcolorbox}

\usepackage{xcolor}          

\definecolor{lightblue}{RGB}{0, 0, 100}

\newtcolorbox{MyBox}{
  colback=white,
  colframe=lightblue,
  fonttitle=\bfseries,
  coltitle=black,
  sharp corners,
  boxrule=1pt,
  left=5pt,
  right=5pt,
  top=5pt,
  bottom=5pt,
  breakable
}

\definecolor{purplish}{HTML}{D8DFE3}
\definecolor{purplishlight}{HTML}{EBEFF3}
\definecolor{purplishdark}{HTML}{FF7F50}

\newtcolorbox[auto counter,number within=section]{rqbox}[2]{
    nameref=#1,
    title=\small{#1}, 
    enhanced,
    attach boxed title to top left={yshift=-6pt, xshift=8pt},
    boxed title style={size=small,boxsep=1pt},
    colframe=purplishdark,colback=white,colbacktitle=purplishdark,
    boxsep=2pt,left=2pt,right=2pt,top=6pt,bottom=2pt,middle=2pt
}

\begin{document}

\title{Tether: A Personalized Support Assistant for Software Engineers with ADHD}

\author{

\IEEEauthorblockN{Aarsh Shah}
\IEEEauthorblockA{University of Calgary\\
Calgary, AB, Canada \\
aarsh.shah@ucalgary.ca} 
\and

\IEEEauthorblockN{Cleyton Magalhaes}
\IEEEauthorblockA{UFRPE\\
 Recife, PE, Brazil \\
cleyton.vanut@ufrpe.br}
\and

\IEEEauthorblockN{Kiev Gama}
\IEEEauthorblockA{CIn-UFPE\\
 Recife, PE, Brazil \\
kiev@cin.ufpe.br}
\and

\IEEEauthorblockN{Ronnie de Souza Santos}
\IEEEauthorblockA{University of Calgary\\
Calgary, AB, Canada \\
ronnie.desouzasantos@ucalgary.ca} 

}


\IEEEtitleabstractindextext{%
\begin{abstract}

\end{abstract}

\begin{IEEEkeywords}

\end{IEEEkeywords}}

\maketitle


\IEEEpeerreviewmaketitle

\begin{abstract}
Equity, diversity, and inclusion in software engineering often overlook neurodiversity, particularly the experiences of developers with Attention Deficit Hyperactivity Disorder (ADHD). Despite the growing awareness about that population in SE, few tools are designed to support their cognitive challenges (e.g., sustained attention, task initiation, self-regulation) within development workflows. We present Tether, an LLM-powered desktop application designed to support software engineers with ADHD by delivering adaptive, context-aware assistance. Drawing from engineering research methodology, Tether combines local activity monitoring, retrieval-augmented generation (RAG), and gamification to offer real-time focus support and personalized dialogue. The system integrates operating system level system tracking to prompt engagement and its chatbot leverages ADHD-specific resources to offer relevant responses. Preliminary validation through self-use revealed improved contextual accuracy following iterative prompt refinements and RAG enhancements. Tether differentiates itself from generic tools by being adaptable and aligned with software-specific workflows and ADHD-related challenges. While not yet evaluated by target users, this work lays the foundation for future neurodiversity-aware tools in SE and highlights the potential of LLMs as personalized support systems for underrepresented cognitive needs.
\end{abstract}

\begin{IEEEkeywords}
ADHD, Assistive Tools, LLMs
\end{IEEEkeywords}

\section{Introduction}
\label{sec:introduction}
Equity, diversity, and inclusion (EDI) are increasingly recognized as critical in software engineering (SE), yet the field continues to lack representation across multiple identity dimensions \cite{santos2024exploring, marquez2024inclusion}. Although diversity is associated with innovation and improved team performance, the SE workforce remains predominantly homogeneous \cite{da2025felt, verma2025differences}. Most EDI-focused research in SE is gender-centered, with much less attention to race, nationality, age, or disability \cite{rodriguez2021perceived, albusays2021diversity, silveira2019systematic, marquez2024inclusion}. This limited scope contributes to inequities in hiring, participation, and in the design of tools and systems that often overlook the needs of diverse users \cite{albusays2021diversity, da2025felt, newman2025get}. Structural barriers— including implicit bias, intersectionality, and cultural taxation—further marginalize underrepresented developers and lead to higher attrition rates \cite{albusays2021diversity, santos2024exploring, da2025felt}. Addressing these challenges requires expanding the focus of EDI research and adopting inclusive practices that reflect the global population \cite{marquez2024inclusion, verma2025differences}.

Building on broader concerns around EDI in SE, recent studies have drawn attention to the specific challenges encountered by developers with ADHD, who represent over 10\% of the industry \cite{stackoverflow2022, gama2025socio}. Despite their significant presence, they often face persistent obstacles related to sustained attention, task initiation and completion, time management, and executive functioning \cite{da2025felt, liebel2024challenges, gama2025socio, morris2015understanding, gama2025socio}. These difficulties are compounded by workplace norms such as synchronous communication, constant context switching, and open office environments, which create additional cognitive strain \cite{newman2025get, morris2015understanding, gama2025socio, da2025felt}. While many rely on coping mechanisms like timeboxing, external reminders, and rigid task structuring, these are often insufficient without broader organizational support \cite{gama2025socio, morris2015understanding, gama2025socio, da2025felt}. Formal accommodations remain underutilized due to stigma, lack of awareness, or unclear processes \cite{gama2025socio, da2025felt, santos2024exploring}. As a result, ADHD developers frequently report underperformance, stress, and job dissatisfaction \cite{gama2025socio, da2025felt}, reflecting a broader neglect of neurodiversity in SE and reinforcing the need for targeted research and inclusive practices \cite{gama2025socio, santos2024exploring, newman2025get}.

In response to these gaps, recent studies across clinical and human-computer interaction domains have begun exploring the potential of Large Language Models (LLMs) as assistive tools for individuals with ADHD, offering novel approaches to address long-standing challenges in attention regulation and emotional support \cite{berrezueta2024future, berrezueta2025integrating, bannett2025applying, olsen2016adhd}. For example, LLM-powered conversational systems have been shown to improve user focus, emotional regulation, and overall engagement through interactive virtual characters tailored to ADHD users’ cognitive and emotional needs \cite{li2025design, berrezueta2024future, carik2025exploring}. These systems incorporate features such as proactive conversation initiation, role-switching, and adaptive feedback to maintain user attention and reduce the cognitive burden of task initiation \cite{li2025design, wang2025artificial, carik2025exploring}. Other investigations into LLM-based interventions, such as those employing ChatGPT for therapy enhancement, report promising results in terms of empathy, communication adaptability, and expanded access to support services, particularly in under-resourced settings \cite{berrezueta2024future, berrezueta2025integrating, bannett2025applying, olsen2016adhd}.

Considering that \textbf{existing LLM applications for ADHD support have primarily targeted general therapy or daily life assistance}, there remains a gap in addressing the specific cognitive demands faced by neurodivergent professionals in high-focus domains such as software development \cite{li2025design, berrezueta2025integrating, bannett2025applying}. This gap leaves unaddressed the challenges of managing focus and maintaining task flow in technical work environments. To explore this problem, we propose the following research question: \textbf{\textit{RQ. How can an LLM-powered tool support software engineers with ADHD in structuring their workflows and maintaining focus during development tasks?}}. This paper presents preliminary results of the design and evaluation of our novel approach, offering a new direction for assistive technologies that better reflect the lived realities of neurodivergent professionals in SE \cite{liebel2024challenges, gama2025socio}.



\section{Background} \label{sec:background}
A growing number of digital tools have been developed to support focus and attention in individuals with ADHD, offering a foundation for technology-driven interventions. These tools address core challenges such as distractibility, poor sustained attention, and difficulty managing tasks \cite{kyriakaki2023mobile, doulou2025managing, puasuarelu2020attention}, while also supporting executive functioning, academic development, emotional regulation, and motivation \cite{powell2017adhd, hernandez2025decade}. 
Clinical research \cite{fleming2012developmental} highlights that some of the effective interventions for ADHD at an adult age consist of externalizing higher-level executive processes (e.g., planning, organization, inhibition, time management) by embedding support into the environment rather than relying solely on internal control. This principle is echoed in evidence-based interventions that leverage structured routines, reminders, and contextual scaffolding to reduce cognitive load and support behavior regulation.

Educational apps like Say-it and Learn use music, animation, and real-time feedback to teach literacy and math through interactive methods \cite{doulou2022mobile, butt2020say}, and Supangan offers gamified lessons with audio feedback to reinforce participation and concept retention \cite{doulou2025managing}. For self-regulation and routines, tools such as TangiPlan guide morning activities, ChillFish promotes calm breathing through biofeedback, and BlurtLine provides tactile feedback to reduce impulsive speech \cite{doulou2022mobile}. Cognitive training is supported by the ADHD Trainer (Tajima Cognitive Treatment), which targets attention and memory with measurable improvements in task performance \cite{carvalho2023evaluation, doulou2022mobile}.

A subset of these tools specifically focuses on attention management and distraction reduction, which are central concerns for individuals with ADHD. Apps like Stayfocusd and Leechblock restrict access to distracting websites, while SimplyNoise uses ambient sound to support concentration \cite{doulou2022mobile}. The N-back app provides cognitive training for working memory and sustained attention \cite{doulou2022mobile}. Living Smart assists adults in structuring routines and reducing disorganization, and Snappy uses motion data to help users monitor impulsivity and attention \cite{powell2017adhd}. CASTT is designed to detect when learners lose attention and provide prompts to help them refocus in classroom settings \cite{doulou2022mobile}. Altogether, these eight applications (i.e., Stayfocusd, Leechblock, SimplyNoise, ADHD Trainer, Living Smart, N-back, CASTT, and Snappy) highlight the range of approaches used to enhance attention regulation and focus among individuals with ADHD across different age groups and contexts.

\section{Method} \label{sec:method}
Following ADHD treatment principles catalogued by Fleming and McMahon \cite{fleming2012developmental}, Tether aims to externalize key executive processes (e.g., sustained attention, task initiation, and planning) through LLM-driven scaffolding. By embedding these supports into the developer’s real-time work environment, the tool helps mitigate the “double-deficit” in self-regulation characteristic of emerging adults with ADHD.  This study adopts the Engineering Research methodology~\cite{ralph2020empirical}, a scientific method that focuses on the design and evaluation of software artifacts to address real-world challenges. Our artifact is a conversational desktop assistant intended to support software engineers with ADHD in managing focus and task engagement. The methodology emphasizes iterative development, user-centered evaluation, and transparency of design decisions to support reproducibility.

\subsection{Tool Development}
Tether was developed as a desktop application to ensure privacy-preserving local execution, minimize disruptions, and offer real-time, adaptive support for ADHD-related challenges, drawing on prior work in software and assistive technologies for ADHD \cite{doulou2022mobile, doulou2025managing, kyriakaki2023mobile, powell2017adhd, puasuarelu2020attention, carvalho2023evaluation}. Its architecture is shaped by four key components: a) a \textbf{bot interface} that delivers natural, adaptive interactions through a conversational assistant, helping reduce cognitive overhead and improve engagement for users with attentional difficulties \cite{santhanam2022botsinse, wessel2022benefits}; b) \textbf{LLM support}, which enables context-aware emotional and technical support without requiring structured input \cite{fan2023large, lin2025can}; c) a \textbf{ retrieval-augmented generation (RAG)} pipeline built with LangChain, enhancing factual grounding and personalization by indexing ADHD-related literature and prior user interactions \cite{gao2023retrieval, dong2025understand}; and d) a lightweight \textbf{gamification engine} that tracks focus-related behaviors and rewards users with points, badges, and interface customizations to reinforce engagement in a non-disruptive way \cite{doulou2025managing, hernandez2025decade, puasuarelu2020attention}. Tether also integrates a local monitoring engine that collects data on active window usage, idle time, and recovery patterns. These signals are used to generate structured prompts that trigger chat interactions and notifications. All components, including the Flask API backend, the local SQLite chat history, and the notification engine, are designed to operate locally, except for external LLM calls.

\subsection{Preliminary Evaluation}
In this preliminary stage, the evaluation was conducted based on a comparison with existing tools for focus and attention support described in the literature (see Section~\ref{sec:background}). This comparison considered the design scope, interaction models, and contextual adaptability of each tool, guided by common ADHD-related challenges such as task initiation and attention regulation. No formal user study was conducted during this stage. This validation strategy aligns with engineering research practices that emphasize early, utility-oriented iteration as part of artifact development~\cite{ralph2020empirical}.

\section{Preliminary Results} \label{sec:findings}
\begin{figure}[htbp]
  \centering
  \includegraphics[width=0.7\linewidth]{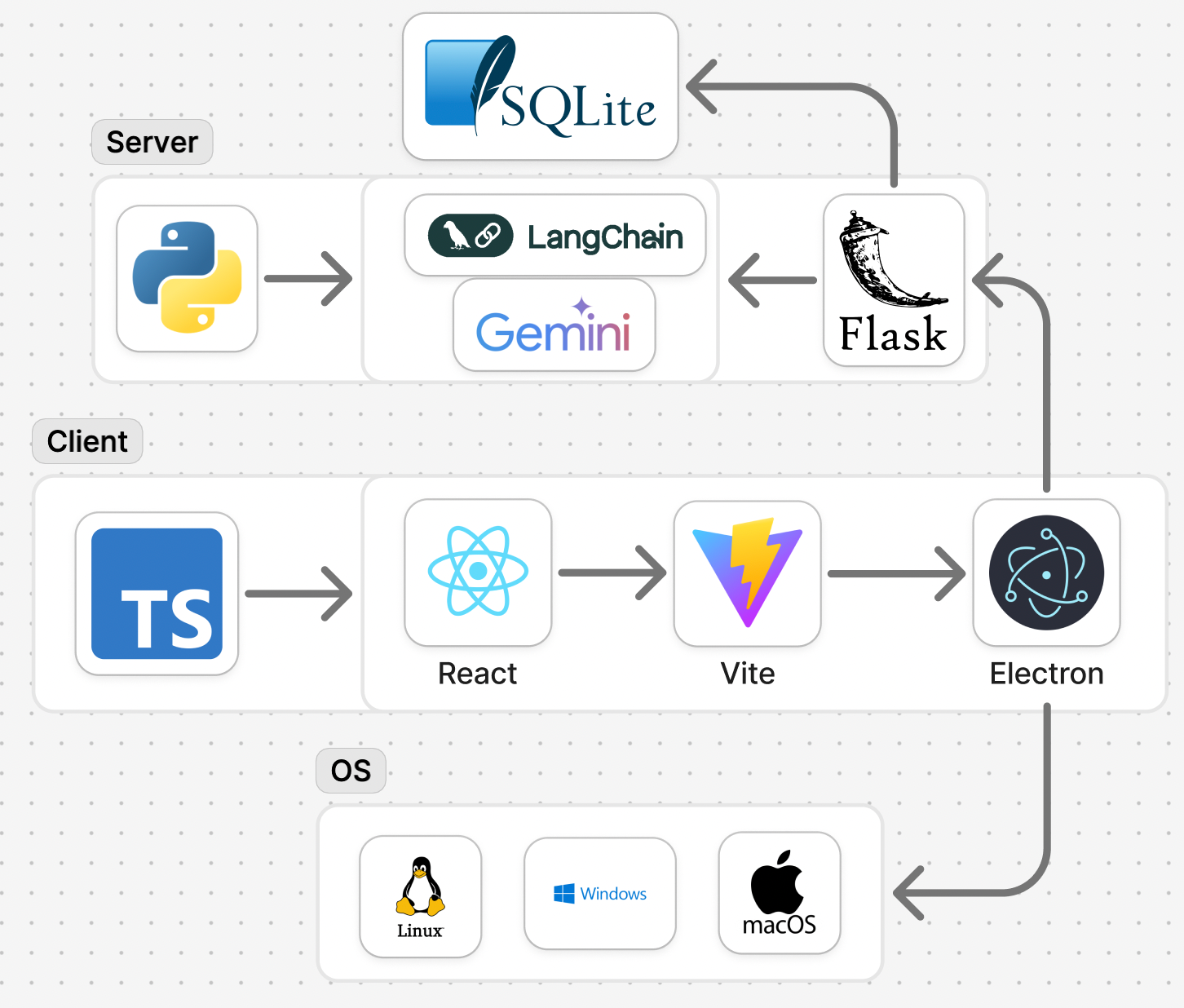}
  \caption{System architecture of the Tether application.}
  \label{fig:tether-architecture}
\end{figure}

Tether is a desktop application developed to support software engineers with ADHD by combining local activity monitoring, contextual prompting, and structured feedback mechanisms. The application is built using a modular architecture consisting of a frontend interface built with Electron and React, and a backend served through a Flask REST API as seen in Figure 1, and it is available at: \url{https://github.com/SeallLab/Tether/releases/tag/v0.3.0}. A local SQLite database is used to store chat history. The system architecture is designed to run entirely on the user’s machine to preserve privacy, with the exception of calls made to any LLMs. Tether includes integrations with native operating system services, including active window tracking, idle state detection, and system notifications. These services continuously record usage patterns and behavioral signals, which are used as context inputs for other components of the system. A retrieval-augmented generation pipeline, implemented using LangChain, indexes ADHD-specific reference materials and previous user interactions to support contextual response generation. The system uses Gemini for language generation and Gemini’s embedding model for document retrieval. Figure~\ref{fig:tether-architecture} provides a high-level overview of the system architecture. Due to the conference page limit, additional implementation details are available here: \url{https://figshare.com/s/9f5e8e201fdf0b26fc36}.

\vspace{-0.5em}
\subsection{User Workflow}
\begin{figure}[htbp]
  \centering
  \includegraphics[width=0.65\linewidth]{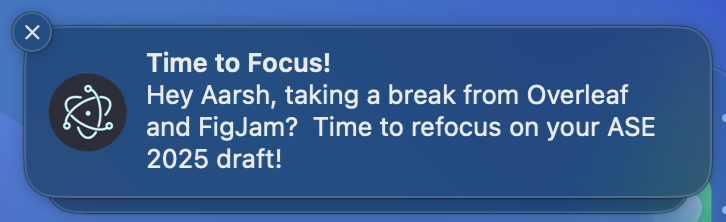}
  \caption{Sample notification for prolonged loss in focus.}
  \label{fig:notification}
\end{figure}

Tether is designed to help software engineers with ADHD stay focused and re-engage when attention drifts. When users step away from their work or stop interacting with their computer for a prolonged period, the tool detects this inactivity and sends a supportive notification. These gentle nudges are tailored based on recent activity, such as which apps were in use or how long the user has been idle. Rather than interrupting users abruptly, Tether delivers encouraging, personalized messages that help them return to their tasks without pressure (see Figure~\ref{fig:notification}).

\begin{figure}[htbp]
  \centering
  \includegraphics[width=1\linewidth]{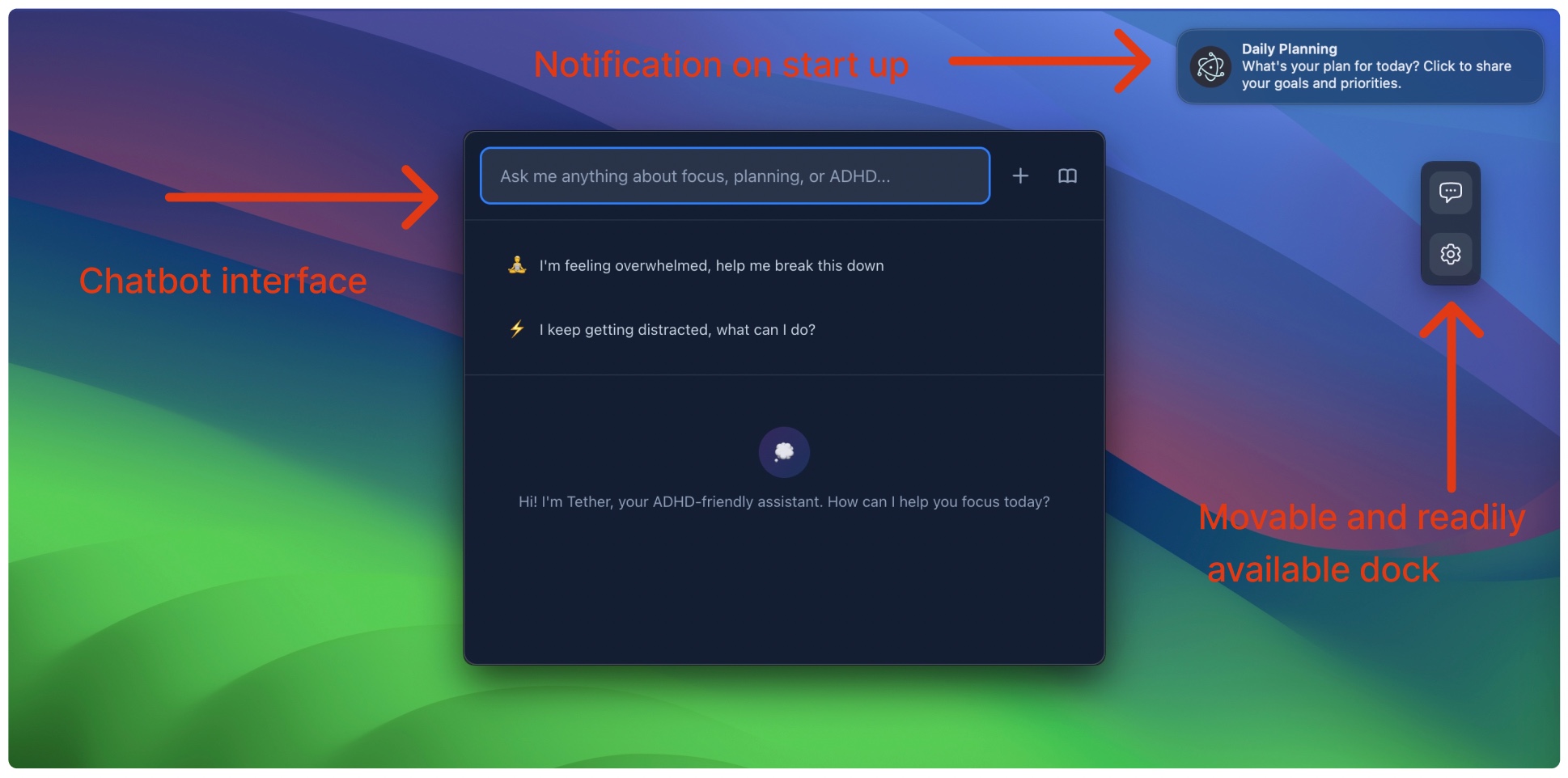}
  \caption{Main user interface and chatbot elements.}
  \label{fig:userinterface}
\end{figure}

In addition to notifications, users can interact with Tether through a built-in chatbot designed to support both emotional regulation and task management. For example, if a user is feeling overwhelmed or stuck on a coding task, they can ask the chatbot for help. The chatbot offers reflective prompts, grounding techniques, and software task guidance based on both recent activity and ADHD-specific reference material. It can suggest small, manageable steps, provide time boxing strategies, or simply offer empathetic conversation. This makes Tether a focus assistant and also a companion that understands the cognitive and emotional challenges (see Figure~\ref{fig:userinterface}).

\begin{figure}[htbp]
  \centering
  \includegraphics[width=0.9\linewidth]{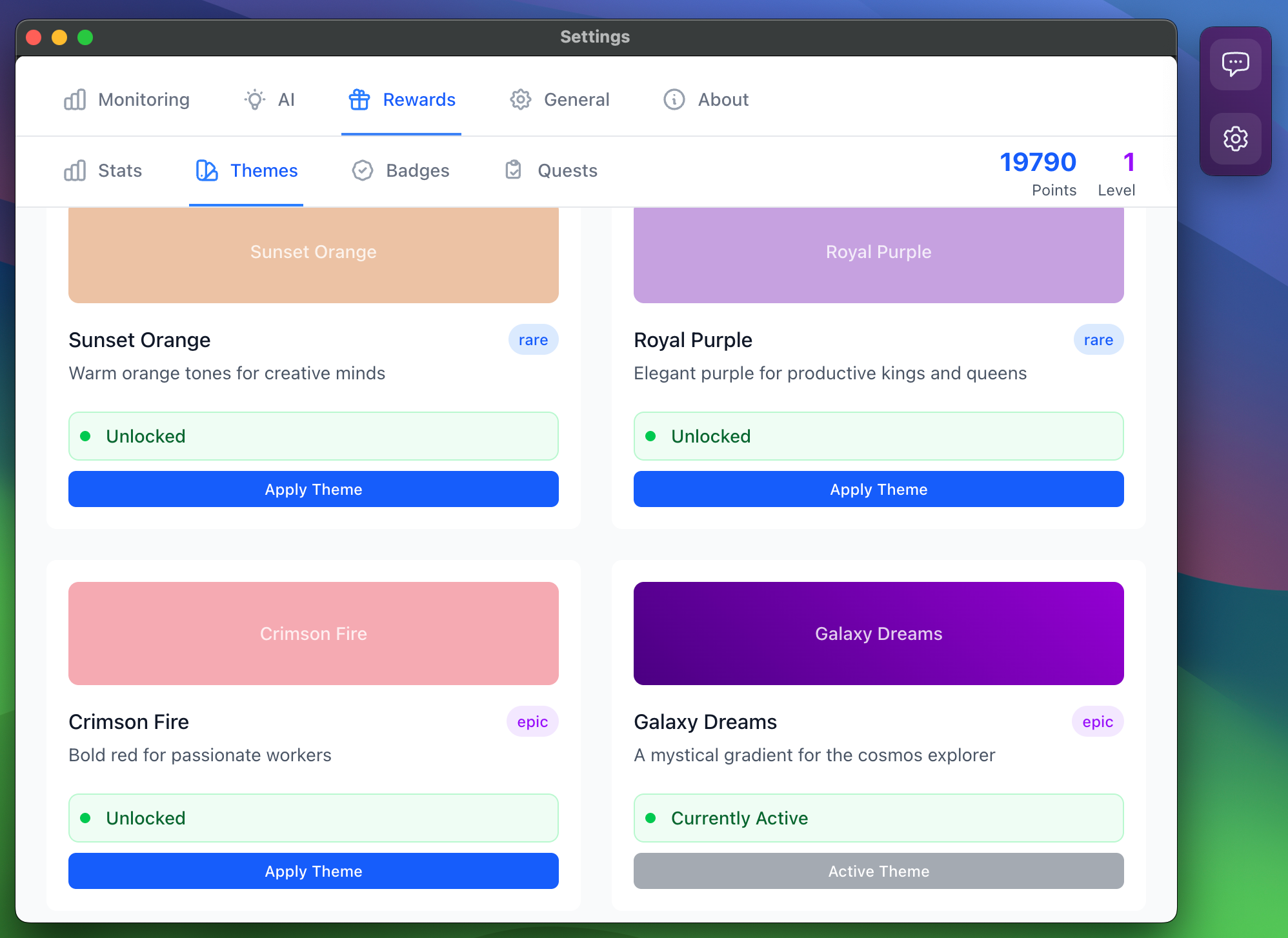}
  \caption{Gamification aspects.}
  \label{fig:gamification}
\end{figure}

To further support motivation and sustained engagement, a gamified progress tracker is included. As users stay focused, switch tasks efficiently, or recover quickly from distractions, they earn points, badges, and milestone rewards. These achievements are displayed in a dedicated interface and can be used to unlock alternate user interface themes. When users hit focus goals or complete challenges, Tether provides real-time positive reinforcement through notifications. This gamification layer is designed to reward effort and celebrate small wins, key motivators for users with ADHD (see Figure~\ref{fig:gamification}).

\vspace{-1em}
\subsection{Internal Process}
Tether’s backend processes are initiated either through passive detection (e.g., user inactivity) or active engagement (e.g., user interacting with the chatbot). The frontend tracks desktop activity, such as mouse movement, keystrokes, and app usage, and sends this data to a Flask backend, which analyzes it for signs of disengagement or context loss. If a trigger is detected, a structured prompt is generated using summaries of recent activity, chat history, and metadata. This prompt follows a modular template that incorporates ADHD-informed principles and guides the LLM's response type (e.g., motivational nudge, task suggestion, or emotional check-in). The system uses a retrieval-augmented generation (RAG) setup via LangChain, indexing both ADHD literature (e.g., the systematic literature reviews cited in Section 2) and user-specific data. The fully constructed prompt is then sent to Gemini’s LLM API. Depending on the context and prompt instructions, the generated response is routed either to the native OS notification service (for nudges) or to the chatbot interface (for conversational support). The result is a responsive, personalized assistant grounded in real-time behavior and therapeutic strategies.

\subsection{Preliminary Evaluation Results}
As a preliminary evaluation, we compared Tether with tools available in the literature that were developed to support individuals with ADHD across behavioral and cognitive domains \cite{doulou2022mobile, doulou2025managing, kyriakaki2023mobile, powell2017adhd, carvalho2023evaluation}. The comparison focused on differences in technology, domain, and focus area. In terms of technology, most existing tools are implemented as browser extensions (e.g., Stayfocusd, Leechblock), mobile apps (e.g., ADHD Trainer, N-back), or wearable devices (e.g., ChillFish). In contrast, Tether is a desktop application powered by LLMs that prioritizes local execution. Regarding the domain, prior tools are designed for general audiences or therapeutic support, whereas Tether specifically targets professional software engineers. Other tools focus areas include distraction blocking, cognitive training, and emotional regulation, but none offer support embedded within software development workflows. In addition to these dimensions, we explored whether each tool supports engagement, interaction, and gamification. This feature-level comparison is presented in Table~\ref{tab:tethertools}.

\vspace{-1em}
\begin{table}[h]
\caption{Comparison of Tether and Existing ADHD-Related Tools}
\label{tab:tethertools}
\tiny
\centering
\begin{tabular}{p{1.5cm} p{0.9cm} p{0.9cm} p{1cm} p{0.7cm} p{0.9cm}}
\toprule
\textbf{Tool} & \textbf{Monitoring} & \textbf{Chat} & \textbf{Dev-Aware} & \textbf{RAG} & \textbf{Gamified} \\
\midrule
Stayfocusd & No & No & No & No & No \\
Leechblock & No & No & No & No & No \\
SimplyNoise & No & No & No & No & No \\
ADHD Trainer & No & No & No & No & No \\
N-back & No & No & No & No & No \\
Living Smart & No & No & No & No & No \\
ChillFish & Yes & No & No & No & No \\
TangiPlan & No & No & No & No & No \\
Supangan & No & No & No & No & Yes \\
Say-it \& Learn & No & No & No & No & Yes \\
\textbf{Tether} & Yes & Yes & Yes & Yes & Yes \\
\bottomrule
\end{tabular}
\vspace{0.2cm}
\begin{flushleft}
\end{flushleft}
\end{table}

\vspace{-1em}

\vspace{-0.5em}
\section{Discussion} \label{sec:discussion}
Neurodivergence in software engineering has received growing attention in recent years, yet practical tools designed to support neurodivergent professionals, particularly those with ADHD, remain scarce. Tether tries to address this gap by introducing a context-aware assistant that integrates directly into development workflows, offering real-time, adaptive support for attention management, task initiation, and emotional monitoring. Unlike existing tools focused on general behavior management or cognitive training, Tether responds to work dynamics and uses retrieval-augmented generation to generate timely, ADHD-informed prompts delivered via passive notifications or conversational chatbot interactions. Hence, the implications of this work are twofold. For research, Tether introduces a new direction for studying how assistive technologies can enhance neuroinclusive practices in software engineering, enabling investigations grounded in real-world behavioral data and LLM-driven interaction. For industry practice, our tool offers a deployable, lightweight support mechanism that integrates into existing tooling and rhythms of software development, reducing friction and enabling software engineers with ADHD to engage more fully and sustainably in their work.

\section{Threats to Validity} \label{sec:limitations}
Following the engineering research guidelines \cite{ralph2020empirical}, relevant threats to validity must be acknowledged. This preliminary version of our tool was tested in a simulated setting with predefined scenarios, while it was being developed, which may not fully capture the needs and behaviors of professionals with ADHD in real-world development environments. In this study, our results were not validated with healthcare professionals specializing in ADHD, and no direct testing was conducted with neurodivergent software practitioners. These limitations point to the need for expert validation to assess the tool’s effectiveness and relevance. Still, this work offers an important starting point for integrating neurodiversity-aware support into developer tools.


\section{Future Work} \label{sec:future}
This paper presents early-stage results based on a working prototype of Tether, focusing on its design and internal validation for supporting software developers with ADHD. Future work includes extending the tool with additional sensing channels, such as camera-based gaze detection and microphone input for ambient noise, to better distinguish between focused inactivity and disengagement. We are also exploring ways to improve the chatbot’s task-specific support by tailoring suggestions to coding and testing activities. For real-world validation, we are preparing two studies: one with healthcare professionals to assess therapeutic alignment and another with software engineers with ADHD to evaluate Tether in daily workflows.

\section{Conclusions} 
\label{sec:conclusions}
To conclude, this paper introduced Tether, an early-stage prototype designed to support software engineers with ADHD using context-aware prompts, an adaptive chatbot, and gamified feedback. Preliminary results show promise for providing personalized, low-disruption support in real-world workflows, with potential to enhance neurodiversity inclusion. However, further development and validation are needed.

\ifCLASSOPTIONcaptionsoff
  \newpage
\fi

\balance
\bibliographystyle{IEEEtran}
\bibliography{references}

\end{document}